\documentclass[preprint,showpacs,amsmath,amssymb,aps]{revtex4}
\usepackage{graphicx}
\begin{document}

\title{ Description of the chiral bands in $^{188,190}Os$}

\author{A. A. Raduta$^{a),b)}$ and  C. M. Raduta$^{a)}$ }

\address{$^{a)}$ Department of Theoretical Physics, Institute of Physics and
  Nuclear Engineering, Bucharest, POBox MG6, Romania}

\address{$^{b)}$Academy of Romanian Scientists, 54 Splaiul Independentei, Bucharest 050094, Romania}

\begin{abstract}
 To a phenomenological core described by the Generalized Coherent State Model a set of interacting particles are coupled. Among the particle-core states one identifies a finite set which have the property that the angular momenta carried by the proton and neutron quadrupole bosons and the particles respectively, are mutually orthogonal.  The magnetic properties of such  states are studied. All terms of the model Hamiltonian satisfy the chiral symmetry except for the spin-spin interaction. There are four bands of two quasiparticle-core dipole states type, which exhibit properties which are specific for magnetic twin bands. Application is made for the isotopes $^{188, 190}$Os.
\end{abstract}


\maketitle

\renewcommand{\theequation}{\arabic{equation}}
\setcounter{equation}{0}

 Some of the fundamental properties of nuclear systems may be evidenced through their  interaction with an electromagnetic field. The two components of the field, electric and magnetic, are used to explore the properties of electric and magnetic nature, respectively. At the end of last century the scissors like states \cite{LoIu1,LoIu2,LoIu3} as well as the spin-flip excitations \cite{Zawischa} have been widely treated by various groups. The scissors mode describes the angular oscillation of proton against neutron system and the total strength is proportional to the nuclear deformation squared which reflects the collective character of the excitation \cite{LoIu3,Zawischa}.

In virtue of this feature it was believed that the magnetic collective properties are in general associated with deformed systems. This is not true due to the magnetic dipole bands, where the ratio between the moment of inertia and the B(E2) value for exciting the first $2^+$ from the ground state $0^+$,
${\cal I}^{(2)}/B(E2)$,
takes large values, of the order of 100(eb)$^{-2}MeV^{-1}$. These large values can be justified
by a large transverse magnetic dipole moment which induces dipole magnetic transitions, but almost no charge quadrupole moment \cite{Frau}. Indeed, there are several experimental data showing that the dipole bands have large values for $B(M1)\sim 3-6\mu^2_N$, and very small values of $B(E2)\sim 0.1(eb)^2$ (see for example Ref.\cite{Jenkins}). The states are different from the scissors mode, they being rather of a shears character. A system with a large transverse magnetic dipole moment may consist of a triaxial core to which a proton prolate  and a  neutron oblate hole orbital are coupled. The maximal transverse dipole momentum is achieved, for example, when $\bf{j}_p$ is oriented along the small axis of the core, $\bf{j}_n$ along the long axis and the core rotates around the intermediate axis.   Suppose the three orthogonal angular momenta form a right trihedral frame. If the Hamiltonian describing the interacting system of protons, neutrons and the triaxial core is invariant to the transformation which changes the orientation of one of the three angular momenta, i.e. the right trihedral frame is transformed to a left type, one says that the system exhibits a chiral symmetry. As always happens, such a symmetry is identified when that is broken and consequently to the two trihedral-s correspond distinct energies, otherwise close to each other. Thus, a signature for a chiral symmetry characterizing a triaxial system is the existence of two $\Delta I=1$ bands which are close in energies.  Increasing the total angular momentum, the gradual alignment of $\bf{j}_p$ and $\bf{j}_n$ to the total $\bf{J}$ takes place and a magnetic band is developed.

Here we attempt another chiral system consisting of one phenomenological core with two components, one for protons and one for neutrons, and two quasiparticles whose total angular momentum 
${\bf J}$ is oriented along the symmetry axis of the core due to the particle-core interaction. In a previous publication we proved that states of total angular momentum ${\bf I}$, where the three components mentioned above carry the angular momenta ${\bf J}_p, {\bf J}_n, {\bf J}$ which are mutually orthogonal, do exist.  Such configuration seems to be  optimal to define  large transverse magnetic moment inducing large M1 transitions. 

The core is described by the Generalized Coherent State Model (GCSM) which is an extension of the Coherent State Model (CSM) for a composite system of protons and neutrons. 
The CSM is based of the ingredients presented below.
The usual procedure used to describe the excitation energies with a given boson Hamiltonian is to diagonalize it and fix the structure coefficients such that some particular energy levels be reproduced. For a given angular momentum, the lowest levels belong to the ground, gamma and beta bands, respectively. For example, the lowest  state of angular momentum 2, i.e. $2^+_1$, is a ground band state, the next lowest, $2^+_2$, is a gamma band state, while $2^+_3$ belongs to the $\beta $ band. The dominant components of the corresponding eigenstates are one, two and three phonon states. The harmonic limit of the model Hamiltonian yields a multiphonon spectrum while by switching on a deforming anharmonicity, the spectrum is a reunion of rotational bands. The correspondence of the two kind of spectra, characterizing the vibrational and rotational regimes respectively, is realized according to the Sheline-Sakai scheme \cite{Sheline}. In the near vibrational limit a certain staggering is observed for the $\gamma$ band, while in the rotational extreme, the staggering is different. The bands are characterized by the quantum number $K$ which for the axially symmetric nuclei is 0 for the ground and $\beta$ bands and equal to 2 for $\gamma$ band. The specific property of a band structure consists of that the E2 transition probabilities within a band is much larger that the ones connecting two different bands. For $\gamma$ stable nuclei, the energies of the states heading the $\gamma$ and $\beta$ bands are ordered as $E_{2^{+}_{\gamma}}> E_{0^{+}_{\beta}}$ while for $\gamma$ unstable nuclei the ordering is reversed. A third class of nuclei  exists where $E_{J^{+}_{\gamma}}\approx E_{J^{+}_{\beta}}$, $J$-even.
{\it These are the fundamental features which should be described by the wave functions of any realistic approach}. CSM builds a restricted basis requiring that the states are orthogonal before and after angular momentum projection and, moreover, accounts for the properties enumerated above. If such a construction is possible, then one attempts to define an effective Hamiltonian which is quasi-diagonal in the selected basis. The CSM is, as a matter of fact, a possible solution in terms of quadrupole bosons.

In contrast to the CSM,  within the GCSM the protons are described by quadrupole
proton-like bosons, $b^{\dagger}_{p\mu}$, while the neutrons by quadrupole neutron-like bosons, $b^{\dagger}_{n\mu}$ .
Since one deals with two quadrupole bosons instead of one, one expects 
to have a more flexible model and to find a simpler solution satisfying the restrictions
required by CSM.  The restricted  collective space is defined  by the states describing the three
major bands, ground, beta and gamma, as well as the band  based on
the  isovector state $1^+$. Orthogonality conditions, required for both intrinsic and projected states, are satisfied by the
following 6 functions which generate, by angular momentum projection,
6 rotational bands:
\begin{eqnarray}
|g;JM\rangle&=&N^{(g)}_JP^J_{M0}\psi_g,~~
|\beta;JM\rangle = N^{(\beta)}_JP^J_{M0}\Omega_{\beta}\psi_g,~~
|\gamma;JM\rangle = N^{(\gamma)}_JP^J_{M2}(b^{\dag}_{n2}-b^{\dag}_{p2})\psi_g,
\nonumber\\
\tilde{|\gamma ;JM\rangle}&=&\tilde{N}^{(\gamma)}_JP^J_{M2}(\Omega^{\dag}_{\gamma,p,2}+\Omega^{\dag}_{\gamma,n,2})\psi_g,~~
|1;JM\rangle = N^{(1)}_JP^J_{M1}(b^{\dag}_nb^{\dag}_p)_{11}\psi_g,\nonumber\\
|{\bar 1};JM\rangle &=& \tilde{N}^{(1)}_JP^J_{M1}(b^{\dag}_{n1}-b^{\dag}_{p1})\Omega^{\dag}_{\beta}\psi_g,
\psi_g = exp[(d_pb^{\dag}_{p0}+d_nb^{\dag}_{n0})-(d_pb_{p0}+d_nb_{n0})]|0\rangle .
\label{figcsm}
\end{eqnarray}
Here, the following notations have been used:
\begin{eqnarray}
\Omega^{\dag}_{\gamma,k,2}&=&(b^{\dag}_kb^{\dag}_k)_{22}+d_k\sqrt{\frac{2}{7}}
b^{\dag}_{k2},~~\Omega^{\dag}_k=(b^{\dag}_kb^{\dag}_k)_0-\sqrt{\frac{1}{5}}d^2_k,~~k=p,n,
\nonumber\\
\Omega^{\dag}_{\beta}&=&\Omega^{\dag}_p+\Omega^{\dag}_n-2\Omega^{\dag}_{pn},~~
\Omega^{\dag}_{pn}=(b^{\dag}_pb^{\dag}_n)_0-\sqrt{\frac{1}{5}}d^2_p,
\nonumber\\
\hat{N}_{pn}&=&\sum_{m}b^{\dag}_{pm}b_{nm},~\hat{N}_{np}=(\hat{N}_{pn})^{\dag},~~
\hat{N}_k=\sum_{m}b^{\dag}_{km}b_{km},~k=p,n.
\label{omegagen}
\end{eqnarray}
Note that a priory we cannot select one of the two sets of states
$\phi^{(\gamma)}_{JM}$ and $\tilde{\phi}^{(\gamma)}_{JM}$ for gamma band, although  one is symmetric and the other asymmetric against the proton-neutron permutation.
The same is true for the two isovector candidates for the dipole states.
In Ref.\cite{Rad3}, results obtained by using alternatively a symmetric and an asymmetric structure
for the gamma band states were presented. Therein it was shown that the asymmetric structure
for the gamma band does not conflict any of the available data. By contrary,
considering for the gamma states an asymmetric structure and fitting the model
Hamiltonian coefficients in the manner described  in Ref.\cite{Rad2}, a better
description for the beta band energies is obtained. Moreover, in that situation
the description of the E2 transition becomes technically very simple. The results obtained in Refs. \cite{Rad2,Rad3} for $^{156}$Gd are relevant in this respect.
For these reasons, here we make the option for a proton-neutron asymmetric
gamma band. All calculations performed so far considered equal deformations for protons and neutrons. The deformation parameter for the composite system is:
\begin{equation}
\rho=\sqrt{2}d_p=\sqrt{2}d_n \equiv \sqrt{2}d.
\end{equation}
The factors $N$ involved in the wave functions are normalization constants calculated in terms of some overlap integrals.

We seek now an effective Hamiltonian for which the projected states (\ref{figcsm}) are, at least in a good approximation, eigenstates in the restricted collective space.
The simplest Hamiltonian fulfilling this condition is:
\begin{eqnarray}
H_{GCSM}&=&A_1(\hat{N}_p+\hat{N}_n)+A_2(\hat{N}_{pn}+\hat{N}_{np})+
\frac{\sqrt{5}}{2}(A_1+A_2)(\Omega^{\dag}_{pn}+\Omega_{np})
\nonumber\\
&&+A_3(\Omega^{\dag}_p\Omega_n+\Omega^{\dag}_n\Omega_p-2\Omega^{\dag}_{pn}
\Omega_{np})+A_4\hat{J}^2,
\label{HGCSM}
\end{eqnarray}
with ${\hat J}$ denoting the proton and neutron total angular momentum.
The Hamiltonian given by Eq.(\ref{HGCSM}) has  only one off-diagonal matrix element in the basis (\ref{figcsm}). That is $\langle\phi^{\beta}_{JM}|H|\tilde{\phi}^{(\gamma)}_{JM}\rangle$.
However, our calculations show that this affects the energies of $\beta$ and $\tilde{\gamma}$ bands by an amount of a few keV. Therefore, the excitation energies of the six bands are in a very good approximation, given by the diagonal element:
\begin{equation}
E^{(k)}_J=\langle\phi^{(k)}_{JM}|H|\phi^{(k)}_{JM}\rangle-
\langle\phi^{(g)}_{00}|H|\phi^{(g)}_{00}\rangle,\;\;k=g,\beta,\gamma,1,\tilde{\gamma},\tilde{1}.
\label{EkJ}
\end{equation}
F spin properties of the model Hamiltonian and analytical behavior of energies and wave functions in the extreme limits of vibrational and rotational regimes have been studied.  Results for the asymptotic regime of deformation suggests that the proposed model generalizes both the two rotor and the two drops models. 

Note that $H_{GCSM}$ is invariant to any p-n permutation and therefore its eigenfunctions have a definite parity. We chose one or another parity for the gamma band, depending on the quality of the overall agreement with the data. We don't exclude the situation when the fitting procedure selects the symmetric $\gamma $ band as the optimal one.
The possibility of having two distinct phases for the collective motion in the gamma band has been considered also in Ref. \cite{Novos} within a different formalism.

The particle-core interacting system is described by the following Hamiltonian:
\begin{eqnarray}
H&=&H_{GCSM}+\sum_{\alpha}\epsilon_{a}c^{\dag}_{\alpha}c_{\alpha}-\frac{G}{4}P^{\dag}P
\nonumber\\
&-&\sum_{\tau =p,n}X^{(\tau)}_{pc}\sum_{m}q_{2m}\left(b^{\dag}_{\tau,-m}+(-)^mb_{\tau m}\right)(-)^m -X_{sS}\vec{J}_F\cdot\vec{J}_c, 
\label{modelH}
\end{eqnarray}
with the notation for the particle quadrupole operator:
\begin{equation}
q_{2m}=\sum_{a,b}Q_{a,b}\left(c^{\dag}_{j_a}c_{j_b}\right)_{2m},~~
Q_{a,b}=\frac{\hat{j}_{a}}{\hat{2}}\langle j_{a}||r^2Y_2||j_b\rangle .
\end{equation}
The core is described by $H_{GCSM}$ while the particle system by  the next two terms standing for a spherical shell model mean-field and pairing interaction of the alike nucleons, respectively. 
The notation $|\alpha\rangle =|nljm\rangle =|a,m\rangle$ is used for the spherical shell model states.
The last two terms, denoted hereafter as $H_{pc}$, express the interaction between the satellite particles and the core through a quadrupole-quadrupole and a spin-spin force, respectively. The angular momenta carried by the core and particles are denoted by $\bf{J}_c (= \bf{J}_{p}+\bf{J}_{n})$ and $\bf{J}_F$, respectively. 
The mean field plus the pairing term is quasi-diagonalized by means of the Bogoliubov-Valatin transformation.
The free quasiparticle term is $\sum_{\alpha}E_{a}a^{\dag}_{\alpha}a_{\alpha}$, while the qQ interaction preserves  the above mentioned form, with the factor $q_{2m}$ changed to:
\begin{eqnarray}
q_{2m}&=& \eta^{(-)}_{ab}\left(a^{\dag}_{j_a}a_{j_b}\right)_{2m}+\xi^{(+)}_{ab}\left((a^{\dag}_{j_a}a^{\dag}_{j_b})_{2m}-(a_{j_a}a_{j_b})_{2m}\right),\;\; \rm{where}
\nonumber\\
\eta^{(-)}_{ab}&=&\frac{1}{2}Q_{ab}\left(U_aU_b-V_aV_b\right),\;\;
\xi^{(+)}_{ab}=\frac{1}{2}Q_{ab}\left(U_aV_b+V_aU_b\right).
\end{eqnarray}
We restrict the single particle space to a proton single-j state where two particles are placed.
In the space of the particle-core states we, therefore, consider the basis defined by:
\begin{equation}
|BCS\rangle\otimes|1;JM\rangle ,
\Psi^{(2qp;J1)}_{JI;M}=N^{(2qp;J1)}_{JI}\sum_{J'}C^{J\;J'\;I}_{J\;1\;J+1}\left(N^{(1)}_{J'}\right)^{-1}\left[(a^{\dag}_ja^{\dag}_j)_J|BCS\rangle\otimes|1;J'\rangle \right]_{IM},
\label{basis}
\end{equation}
where $|BCS\rangle$ denotes the quasiparticle vacuum while $N_{JI}$ is the norm of the projected state. 
The formalism described above was applied for two isotopes $^{188,190}$Os.  In choosing the mentioned isotopes we had in mind their triaxial shape behavior reflected by the signature 
\begin{equation}
E_{2^+_g}+E_{2^+_{\gamma}}=E_{3^{+}_{\gamma}}.
\end{equation}
Indeed, this equation is obeyed with a deviation of 2 keV for $^{188}$Os and 11 keV for $^{190}$Os.
We calculated first the excitation energies for the bands described by the angular momentum projected functions
$|g;JM\rangle\otimes|BCS\rangle,~ |\beta;JM\rangle \otimes|BCS\rangle, |\gamma;JM\rangle \otimes|BCS\rangle, |1;JM\rangle \otimes|BCS\rangle, |{\bar 1};JM\rangle\otimes|BCS\rangle $  and the particle-core Hamiltonian $H$. Several parameters,like the structure coefficients defining  the model Hamiltonian and the deformation parameters $\rho$, are to be fixed.  For a given $\rho$ we determine the parameters involved in $H_{GCSM}$ by fitting the excitation energies in the  ground, $\beta$ and $\gamma$ bands, through a least square procedure. We varied then $\rho$ and kept that value which provides the minimal root mean square of the results deviations from the corresponding experimental data. Excitation energies of the phenomenological magnetic bands  are free of any adjusting parameters. To fix the strengths of  pairing and $Q.Q$ interactions we were guided by  Ref. \cite{Lima}, where spectra of some Pt even-even isotopes  where interpreted  with a particle-core Hamiltonian, the core being described by the CSM. The two quasiparticle energy for the proton orbital $h_{11/2}$  was taken 1.947 MeV for $^{188}$Os and 2.110 meV for $^{190}$Os , these values being close to the ones yielded by a BCS treatment in the extended space of single particle states.
The parameters mentioned above have the values listed in Table I.
\begin{table}[h!]
\scriptsize{
\begin{tabular}{|cccccccccccc|}
\hline
   &$\rho=d\sqrt{2}$  &   $A_1$[keV]   &     $A_2$[keV]    &   $A_3$[keV]    &   $A_4$[keV] &  $X'_{pc}$[keV]  &  $X_{sS}$[keV] &  $g_p$[$\mu_N$]  &  $g_n$[$\mu_N$] &  $g_F$[$\mu_N$] &r.m.s.[keV]
\\
\hline
$^{188}$Os&  2.2   &   438.7   & -93.8   &   -70.5  &   9.1  &  1.02  &  3.0 &0.828 &-0.028&1.289&16.93\\
$^{190}$Os & 2.0   & 366.1     & 92.6  & 24.0    &    12.2& 1.66& 2.0& 0.7915 & 0.0086 &1.289& 18.63\\  
\hline
\end{tabular}}
\caption{\scriptsize {The structure coefficients of the model Hamiltonian were determined  by a least square procedure. On the last column the r.m.s. values characterizing the deviation of the calculated and experimental energies are also given. The deformation parameter $\rho$ is adimensional. The parameter $X'_{pc}$ is related to $X_{pc}$ by:
$X'_{pc}=6.5\eta^{(-)}_{\frac{11}{2}\frac{11}{2}}X_{pc}.$ }}
\end{table}
Excitation energies calculated with these parameters are compared with the corresponding experimental data, in Figs. 1,2. One notes a good agreement of results with the corresponding experimental data.
\begin{figure}[h]
\includegraphics[width=7.5cm]{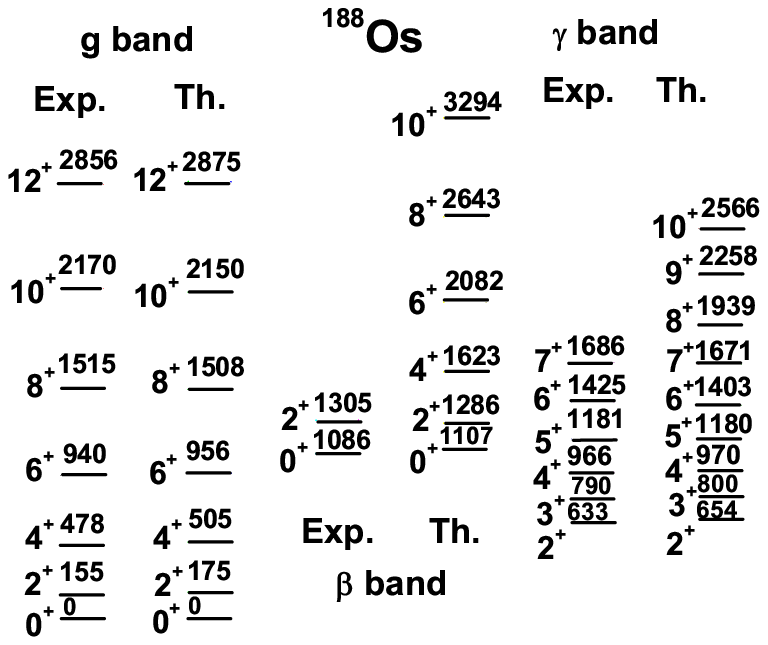}   \includegraphics[width=7.5cm]{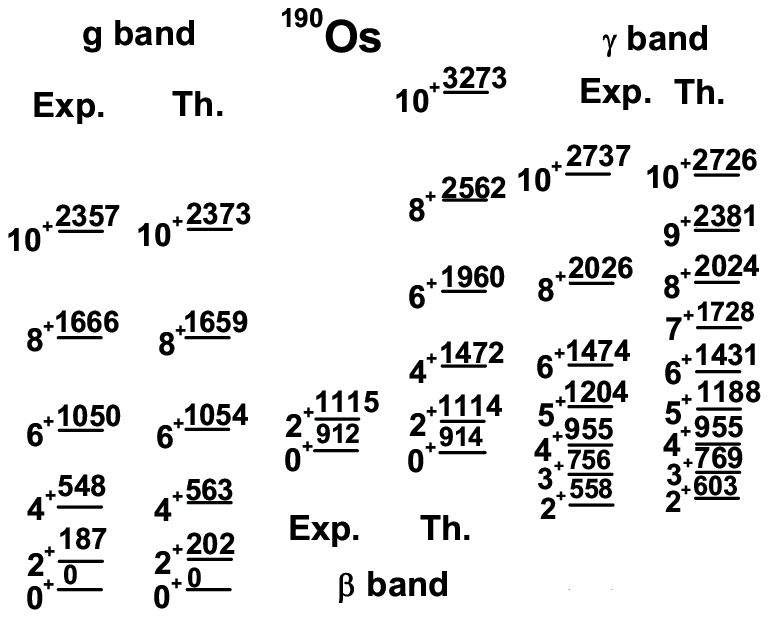}
\begin{minipage}[h]{7cm}
\caption{\scriptsize{Experimental (Exp.) and calculated (Th.) excitation energies in ground, $\beta$ and $\gamma$ bands of $^{188}$Os. Data are taken from \cite{Bal1}. }}
\end{minipage}\ \  
\begin{minipage}[h]{7.5cm}
\caption{\scriptsize{The same as in Fig.1 but for $^{190}$Os with data from Ref.\cite{Bal2}.}}   
\end{minipage}
\end{figure}
\begin{figure}[h!]
\includegraphics[width=7.5cm]{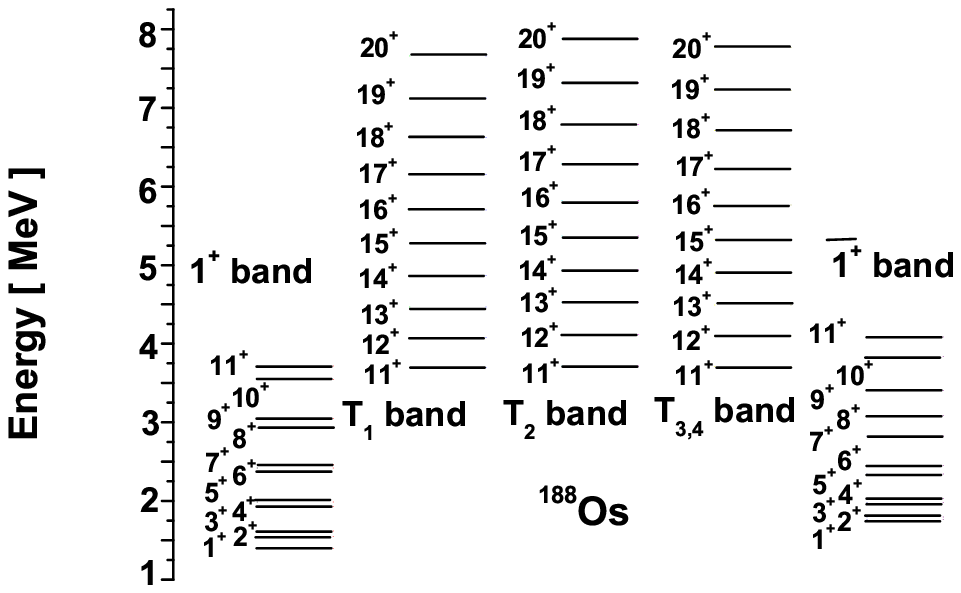} \includegraphics[width=7.5cm]{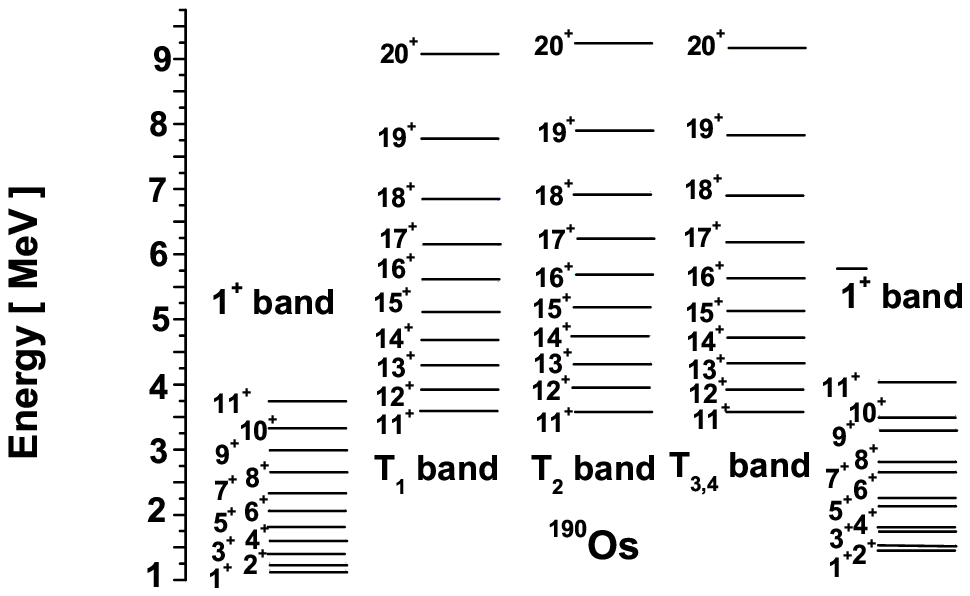}
\begin{minipage}[h]{7.5cm}
\caption{\scriptsize{Excitation energies for the yrast (lower-left) and non-yrast (lower-right) boson dipole states of $^{188}$Os. The twin bands $T_1$ and $T_2$ are also shown.}}    
\end{minipage}\ \
\begin{minipage}[h]{7.5cm}
\caption{\scriptsize{The same as in Fig. 3 but for $^{190}$Os. Here the dipole bands from the lower columns are described by $|1;JM\rangle $ and $|{\bar 1};JM\rangle $.}}
\end{minipage}
\end{figure}
\begin{figure}[h!]
\includegraphics[width=7.5cm]{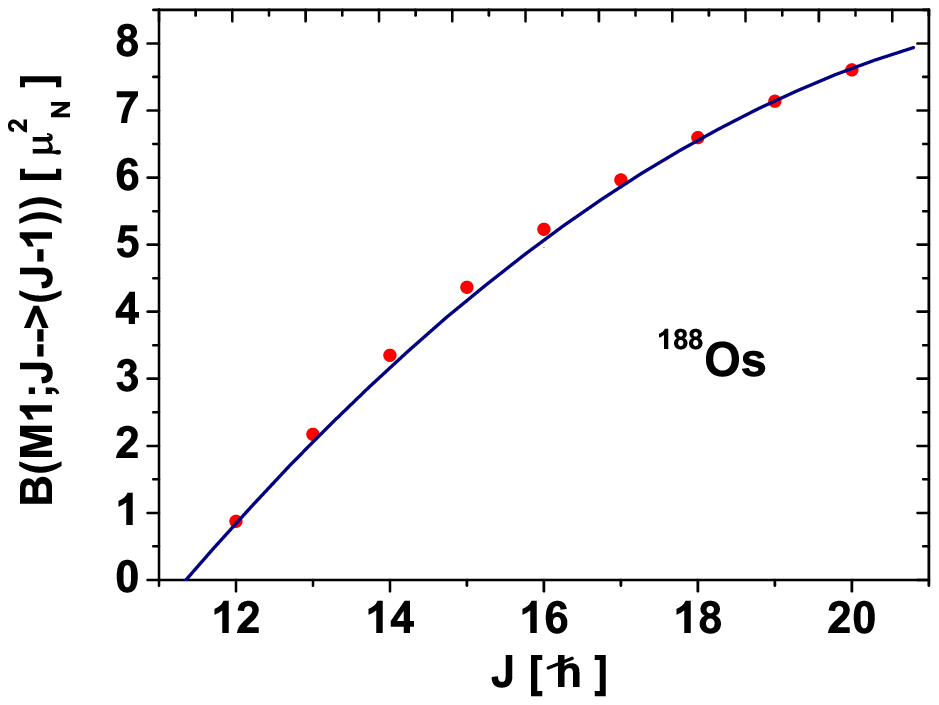} \includegraphics[width=7.5cm]{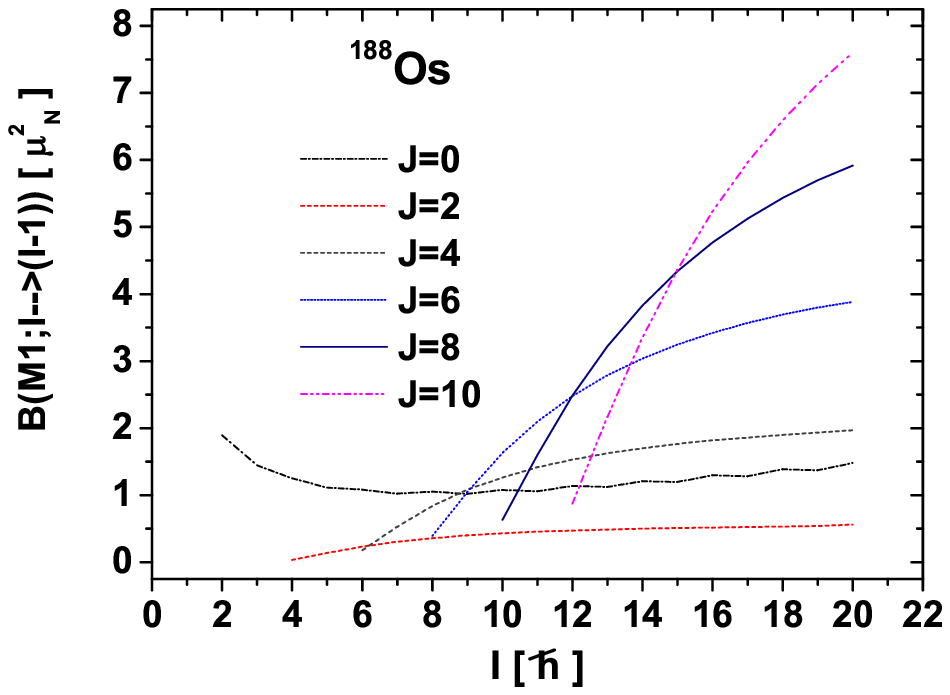}
\begin{minipage}[h]{7.5cm}
\caption{\scriptsize{The BM1 values associated with the dipole magnetic transitions between two consecutive levels in the $T_1$ band of $^{188}$Os. The results are interpolated with a second rank polynomial (full curve). The gyromagnetic factors employed are
$g_p=0.828 \mu_N$, $g_n=-0.028\mu_N$ and $g_F=1,289\mu_N$.}}
\end{minipage}\ \
\begin{minipage}{7.5cm}
\caption{\scriptsize{The magnetic dipole reduced probabilities within the two quasiparticle-core bands  corresponding to the quasiparticle total angular momentum J. The gyromagnetic factors are the same as those used in Fig. 5.}} 
\end{minipage}
\end{figure}
 Unfortunately, there is no available data concerning the magnetic states. However, in Refs.
\cite{Bal1,Bal2} the states of 1304.82 keV and 1115.5 keV  in $^{188}$Os and $^{190}$Os respectively, performs a M1 decay to the ground band states. These states could tentatively be associated to the heading states of the two dipole bands which are located at  1400 and 1538 keV, respectively. For $^{188}$Os, the states $|1;JM\rangle$ are not in a natural order from $J\ge 6$. Indeed,
the yrast states belong to the $1^+$ band except the states with $J=6, 8, 10$ which are of ${\bar 1}^+$ type. Similarly, nonyrast states have a ${\bar 1}^+$ character except the states of
$J=6, 8, 10$, which are of $1^+$ type.

If in the expression of H (\ref{modelH}) one ignores the spin-spin term, the resulting Hamiltonian exhibits a chiral symmetry. The spin-spin interaction however breaks such a symmetry. Indeed, changing alternatively the sign of ${\bf J}_F, {\bf J}_p, {\bf J}_n$ one obtains three distinct interactions which, moreover, are different from the initial one. Associating to each of these interactions a band, one obtains a set of four twin bands among which any two are related by a chiral transformation. In Figs. 3 and 4 the chiral bands $T_1$ and $T_2$ are associated to the actual Hamiltonian given by Eq. (\ref{modelH}) and the one obtained by the chiral transformation  ${\bf J}_F\to -{\bf J}_F$, respectively while the bands $T_3$ and $T_4$ are degenerate and correspond to the transformations ${\bf J}_p\to -{\bf J}_p$ and ${\bf J}_p\to -{\bf J}_p$ respectively. The degeneracy is caused by the fact that in both cases the transformed spin-spin interaction is asymmetric with respect to the p-n permutation and therefore their averages with the two quasiparticle-dipole core states, which are asymmetric, are vanishing. Remarkable the fact that by enlarging the particle-core space with the  $2qp\otimes \Phi^{(g)}_J$ states, then the interaction between the opposite parity $2qp\otimes core$ states due to the spin-spin term, would determine another two bands of mixed symmetry, characterized also by large M1 rates. The description of such bands will be presented elsewhere.  The bands treated here, $T_1, T_2$ and $T_{3,4}$ have, indeed, properties which are specific to chiral bands: i) First of all, as proved in Ref. \cite{AAR2014}, the  trihedral 
$({\bf J}_p, {\bf J}_n, {\bf J}_F)$ is orthogonal for some total angular momenta and almost orthogonal for the rest. Since the involved proton state is $h_{11/2}$ and the fermion angular momentum is $J=10$ with the projection $M=J$, this is aligned to the mentioned axis, which is perpendicular onto the plane of the orthogonal vectors ${\bf J}_p$ and ${\bf J}_n$;
ii) The energy spacings in the two bands have similar behavior as function of the total angular momentum; iii) the staggering function $(E(J)-E(J-1))/2J$ is almost constant;
iv) The most eloquent property is the large B(M1) values for the transition between two consecutive levels. The B(M1) values associated to the intra-band transitions are large despite the fact that the deformation is typical for a transitional spherical-deformed region. This property is shown in Fig, 5. The fact that the large transition matrix elements are associated with a chiral configuration of the involved angular momenta is illustrated in Fig 6 where one sees that large B(M1) values are achieved for large quasiparticle total angular momentum projection on the symmetry axis. The magnetic dipole transition probabilities have been calculated with the transition operator:
\begin{equation}
M_{1,m}=\sqrt{\frac{3}{4\pi}}\left(g_pJ_{p,m}+g_nJ_{n,m}+g_FJ_{F,m}\right).
\end{equation}
Considering for the core's magnetic moment the classical definition, one obtains an analytical expression involving the quadrupole coordinates and their time derivatives of first order, which can be further calculated by means of the Heisenberg equation. Finally, writing the result in terms of quadrupole boson operators and identifying the factors multiplying the proton and neutron angular momenta with the gyromagnetic factors of proton and neutrons, one obtains:
\begin{equation}
\left(\begin{matrix}g_p\cr g_n\end{matrix}\right)=\frac{3ZR_0^2}{8\pi k_p^2}\frac{Mc^2}{(\hbar c)^2}\left(\begin{matrix}A_1+6A_4\cr \frac{1}{5}A_3\end{matrix}\right),
\label{gpgn}
\end{equation}
where $Z$ and $R_0$ denote the nuclear charge and radius, while $M$ and $c$ are the proton mass and light velocity. $k_p$ is a parameter defining the canonical transformation relating the coordinate and conjugate momenta with the quadrupole bosons, while $A_1, A_3, A_4$ are the structure coefficients involved in $H_{GCSM}$. Since within the GCSM the core gyromagnetic factor is
\begin{equation}
g_c=\frac{1}{2}(g_p+g_n),
\end{equation}
we may identify it  with the liquid drop value, $Z/A$,  and consequently the canonicity coefficient acquires the expression:
\begin{equation}
k_p^2=\frac{3}{16\pi}AR_0^2\frac{Mc^2}{(\hbar c)^2}\left(A_1+6A_4+\frac{1}{5}A_3\right).
\end{equation}
Inserting this in Eq.(\ref{gpgn}), the gyromagnetic factors are readily obtained. Their values are listed in Table 1.
The fermion gyromagnetic factor corresponds to the proton orbital h$_{11/2}$ with the spin composing term quenched by a factor 0.75. With this expression for the transition operator we also calculated the B(M1) value for the transitions $1^+\to 0^+_g$ and $1^+\to 2^+_g$. The results are 0.2772$\mu^2_N$, 0.0139 $\mu^2_N$ for $^{188}$Os and 0.1752$\mu^2_N$, 0.0085$\mu^2_N$ for
$^{190}$Os. As a matter of fact this proves that the chiral bands $T_1$ and $T_2$ are of different nature than the low lying scissor mode, they being essentially determined by the moment of inertia dependence on the angular momentum.
\begin{figure}[h!]
\begin{center}
\includegraphics[width=0.5\textwidth]{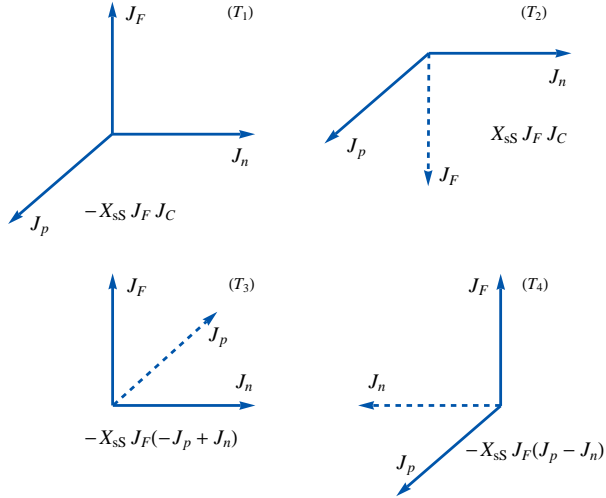}
\end{center}
\caption{ The four frames are related by a chiral transformation. The spin-spin interaction corresponding to each thriedral is also mentioned. They generate the bands $T_{i}$ with i=1,2,3,4, respectively.}
\label{chiralframes}
\end{figure} 
The model used in the present paper was positively tested in a previous publication by  applying it to the case of $^{192}$Pt. Here we used the same ingredients for another two triaxial isotopes, $^{188,190}Os$. By contrast of Ref.\cite{AAR2014} where the gyromagnetic factor of neutrons was taken as 1/5 of $g_p$, here both factors  are calculated in a consistent manner and thus they depend explicitly on the structure coefficients involved in the collective Hamiltonian. Our work proves that the mechanism for chiral symmetry breaking, which also favor a large transversal component for the dipole magnetic transition operator \cite{Frau}, is not unique.

Our description is different from the ones from literature in the following respects. The previous formalisms were focussed mainly on the odd-odd nuclei although few publications refer also to even-odd \cite{Mukhop} and even-even isotopes \cite{Luo}. Our approach concerns the even-even systems and is based on a new concept.
While until now there were only two magnetic bands related by a chiral transformation, here we found four magnetic bands,among which two are degenerate, having this property.
Indeed, consider the thriedral-s $(\bf{J}_p,\bf{J}_n, \bf{J}_F)$, $(\bf{J}_p,\bf{J}_n, -\bf{J}_F)$, $(-\bf{J}_p,\bf{J}_n, \bf{J}_F)$, $(\bf{J}_p,-\bf{J}_n, \bf{J}_F)$ denoted by the same letters as the associated bands, i.e. $T_1$, $T_2$, $T_3$ and $T_4$, respectively. Any pair of these thriedral-s have components related by a chiral transformation. Moreover, they determine four distinct spin-spin interaction terms: $(\bf {J}_F \cdot \bf {J}_c); (-\bf {J}_F \cdot \bf {J}_c); (\bf {J}_F \cdot
(-\bf {J}_p+\bf{J}_n); (\bf {J}_F \cdot(\bf {J}_p-\bf {J}_n)$, each of them affecting the chirally symmetric and degenerate spectrum.  Note that the product of two chiral transformations is a chiral transformation and also that two of the mentioned interactions are symmetric  against the p-n permutations while the other two are asymmetric. As shown in Fig. 7  the pairs  $(T_1, T_2)$, $(T_1, T_3)$ and  $(T_1,T_4)$  components are related by chiral transformations. Also the other pairs, $(T_2,T_3), (T_2,T_4)$ and $(T_3,T_4)$, components are related by the chiral transformations:
${\bf J}_n\to -{\bf J}_n$, ${\bf J}_p\to -{\bf J}_p$ and ${\bf J}_F\to -{\bf J}_F$ respectively. This assertion becomes evident if one compares the corresponding transformed spin-spin Interaction. Concluding, any pair of the four bands are chiral partner bands. The degeneracy of the $T_3$ and $T_4$ bands reflects the pn symmetry of the $X_{sS}=0$ Hamiltonian. This comment explains the appearance of four chiral bands. Note that there are two symmetries broken, that of p-n permutation and the chiral one. These features are suggestively presented in Fig.\ref{chiralframes}.

Here we considered two proton quasiparticle bands but alternatively we could chose two neutron quasiparticles  and one proton plus one neutron quasiparticle bands. Of course, the last mentioned bands would describe an odd-odd system. We already checked that a two neutron quasiparticle band is  characterized by a non-collective M1 transition rate. This feature suggests that, indeed, the orbital magnetic moment carried by protons play an important role in determining a chiral magnetic band. The core is described by angular momentum projected states from a proton and neutron coherent state as well as from its lowest order polynomial 
excitations. Among the three chiral angular momentum components two are associated to the core and one to a two quasiparticle state. By contradistinction, the previous descriptions devoted to odd-odd system, use a different picture. The core carries one angular momentum and, moreover, its shape structure determines the orientation of the other two angular momenta associated to the odd proton and odd neutron, respectively. For odd-odd nuclei several groups identified twin bands in medium \cite{Petrache96,Simon2005,Vaman,Petrache06}  and even heavy mass \cite{Balab} regions.
Theoretical approaches are based mainly on  a triaxial rotor-two quasiparticle coupling, which was earlier formulated and widely used by the group of Faessler \cite{Fas1,Fas2,Fas3,Fas4}.

Experimental data for chiral bands in even-even nuclei are desirable. These would encourage us to extend the present description to a systematic study of the chiral features in even-even nuclei.

{\bf Acknowledgment.} This work was supported by the Romanian Ministry for Education Research Youth and Sport through the CNCSIS program nucleu, ELI, IFIN-HH/2014.

\end{document}